\documentclass[notitlepage,onecolumn,aps,showpacs,preprintnumbers,amsmath,amssymb,superscriptaddress]{revtex4-1}

\usepackage[pdftex]{graphicx}
\usepackage{amsmath}
\usepackage{amsfonts}
\usepackage{amssymb}
\usepackage{color}
\usepackage{float}
\usepackage{color}
\usepackage{natbib}

\begin{document}
\title{Thermodynamic Limits of Energy Harvesting from Outgoing Thermal Radiation}
\bigskip
\author{Siddharth Buddhiraju}
\author{Parthiban Santhanam}
\author{Shanhui Fan}
\affiliation{Ginzton Laboratory, Department of Electrical Engineering, Stanford University, Stanford, California 94305, USA}
\date{\today}
\medskip
\widetext

\begin{abstract}
We derive the thermodynamic limits of harvesting power from the outgoing thermal radiation from the ambient to the cold outer space. The derivations are based on a duality relation between thermal engines that harvest solar radiation and those that harvest outgoing thermal radiation. In particular, we derive the ultimate limit for harvesting outgoing thermal radiation, which is analogous to the Landersberg limit for solar energy harvesting, and show that the ultimate limit far exceeds what was previously thought to be possible.  As an extension of our work, we also derive the ultimate limit of efficiency of thermophotovoltatic systems. 
\end{abstract}
\maketitle
Given any heat flow across a temperature difference, it is possible to construct an engine that extracts work from this flow. The sun, being our primary heat source, supplies a radiative heat flux that sustains both the ambient temperature and the biochemical processes on earth. A significant fraction of the literature on renewable energy conversion has focused on the fundamental limit of harvesting the heat flow from the sun to the earth \citep{shockleyqueisser, henry, yablonovitch, yablonovitch1, yablonovitch3, green, greenauger, greenbook, yupnas, yuprl}. For the thermodynamic analysis in this article, we term this situation ``positive illumination'', where a thermodynamic converter placed on the earth (such as a solar cell), receives a net radiative heat flux and converts a part of it to work.\\

On the other hand, in order for the earth to maintain its temperature, it must also radiate a heat flux that is approximately equal to the incoming heat flux from the sun. Since the sun occupies a very small view factor when looking from the earth's surface, most of the earth's emission is radiated to the cold outer space. The heat flow from the earth to the outer space can also be utilized towards renewable energy harvesting. We call this latter situation ``negative illumination'', where the thermodynamic converter placed on earth seeks to convert a part of the outgoing radiative flux to work. \\

In recent years, there has been an increasing interest in utilizing outgoing radiative flux for renewable energy applications. Substantial progress has been made in radiative cooling, which seeks to lower the temperature of an object to below the ambient air temperature passively \citep{catalanottiRC, berdahlRC, gentleRC, rephaeliRC, ramanRC, nanfangRC,  smithRC, hossainRC, chenRC, zhaiRC, minnichRC, goldsteinRC} by radiating heat away to the outer space. With radiative cooling achieved, it is possible to arrange for a heat engine to operate between the ambient and such a cooled object to extract work. Thus, motivated by the experimental developments of radiative cooling, there have been several theoretical analyses on the amount of work that can in principle be extracted from outgoing radiative flux \citep{berdahl, capasso, strandberg1, strandberg2, parthi, gangchen}. \\

In this paper, we provide a discussion of the fundamental thermodynamic limits of the amount of energy that can be harvested from outgoing radiative flux. We show that the ultimate limit can be derived in a fashion analogous to the derivation of the Landsberg limit in solar energy conversion. This ultimate limit results in a power density of extractable work that far exceeds what was previously thought to be possible for harvesting outgoing thermal radiation. In deriving these limits, we also note an interesting duality relation between the cases of positive and negative illumination. As a result of this duality, every major fundamental limit in the positive illumination case for harvesting incoming solar radiation has a corresponding limit in the negative illumination case for harvesting outgoing thermal radiation. \\

\begin{figure}
\centering
\includegraphics[scale=0.47]{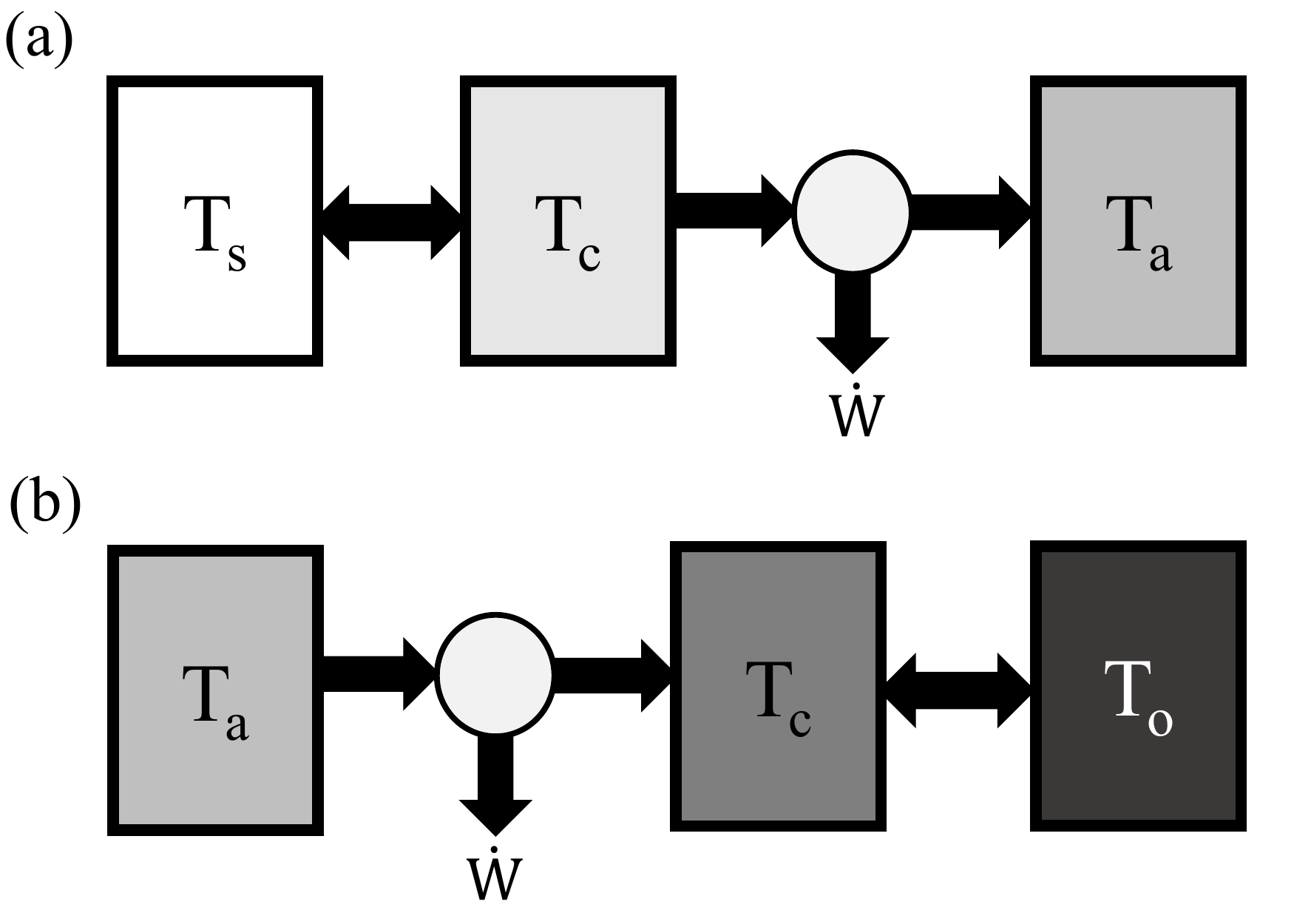}
\caption{Setups to obtain the blackbody limits of energy harvesting under (a) positive illumination, and (b) negative illumination. The temperatures of the sun, the ambient and the outer space respectively are $T_s$, $T_a$ and $T_o$. The rectangles with black outlines represent blackbodies at the indicated temperatures, and are shaded from white (hot) to dark grey (cold). The black arrows indicate the direction of heat flux. The circular disks represent Carnot engines. The intermediate blackbody absorber/emitter is at a steady state temperature $T_c$, and work extracted by the Carnot engine is obtained by setting the net heat flux gained by this intermediate blackbody to zero.}
\label{SingleEmitterFigure}
\end{figure}

This article is organized as follows. In Section 1, we briefly discuss the known results of blackbody limits of positive and negative illumination to lay the foundation for the rest of the paper. In Section 2, we discuss the multi-color limits of positive and negative illumination. In Section 3, we discuss the Landsberg limit, which is the ultimate theoretical limit for harvesting incoming solar radiation, and extend it to the case of negative illumination to obtain the ultimate limit of harvesting outgoing thermal radiation. In Section 4, as an extension of our theoretical analysis, we provide a derivation of the ultimate limit of efficiency of thermophotovoltatic systems.  We conclude in Section 5.

\section{Blackbody Limit}
In order to illustrate the correspondence between the various fundamental limits under positive and negative illumination, we start with a brief discussion of a well-known result in solar energy conversion concerning the limiting efficiency of solar thermophotovoltaic systems, and its connection to a limit on energy harvesting from the earth's thermal emissions as derived in Ref. \citep{capasso}. In this article, we shall follow the convention of Ref. \citep{greenbook} in referring to various limits. The limits considered for this section, for example, will therefore be called the blackbody limits. \\

The limiting efficiency of solar thermophotovoltaic systems is derived \citep{devosBook, greenbook} using the setup shown in Fig. \ref{SingleEmitterFigure}(a). In this setup, an intermediate blackbody at a steady state temperature $T_c$ receives a net radiative heat flux from a source (such as the sun) at $T_s$, and converts a part of this heat to work through a Carnot engine that uses the blackbody as the heat source and the ambient at the temperature $T_a$ as the heat sink. The work extracted by the Carnot engine is $\dot{W} = \sigma(T_s^4-T_c^4)(1-T_a/T_c)$, where the first factor is the the heat input to the blackbody from the radiative exchange between it and the heat source, and the second factor is the Carnot efficiency. Here, $\sigma$ is the Stefan-Boltzmann constant. Assuming $T_s = 6000$ K corresponding approximately to the temperature of the sun, and an ambient temperature of $T_a = 300$ K, the system attains a maximum efficiency of $\eta = \dot{W}/\sigma T_s^4= 85.4\%$ when the intermediate blackbody is at the optimal temperature of $T_c = 2544$ K.\\

A setup related to Fig. \ref{SingleEmitterFigure}(a), shown in Fig. \ref{SingleEmitterFigure}(b), was used to derive a limit on harvesting outgoing thermal radiation in Ref. \citep{capasso}. In this setup, an intermediate blackbody emitter facing the cold outer space at temperature $T_o$ lowers its temperature $T_c$ to below the ambient temperature $T_a$ through the radiative cooling process. A Carnot engine can then operate between the ambient and the intermediate blackbody to extract work. The work extracted by the Carnot engine is $\dot{W}=(T_a/T_c-1)\cdot \sigma(T_c^4-T_o^4)$, where the second factor is the net radiative flux emitted by the intermediate blackbody to the outer space, and the first factor arises from the relation between the extracted work and the rejected heat to the sink by a Carnot engine. For the ambient temperature $T_a = 300$ K and the cold outer space taken to be at $T_o = 3$ K, the work extracted is maximized to 48.4 $\mathrm{W/m^2}$ when $T_c = 225.1$ K. \\

Importantly, the positive and negative illumination setups of Figs. \ref{SingleEmitterFigure}(a) and \ref{SingleEmitterFigure}(b) exhibit a striking duality relation in the arrangement of their components. The two setups can be mapped to each other by interchanging the Carnot engine and the radiative heat transfer element, with appropriate changes to the temperatures of the reservoirs. In the following sections, we shall exploit this duality relation to derive negative illumination limits that can significantly exceed the blackbody limit. 
\section{Multi-Color Limit}
\begin{figure}
\centering
\includegraphics[scale=0.34]{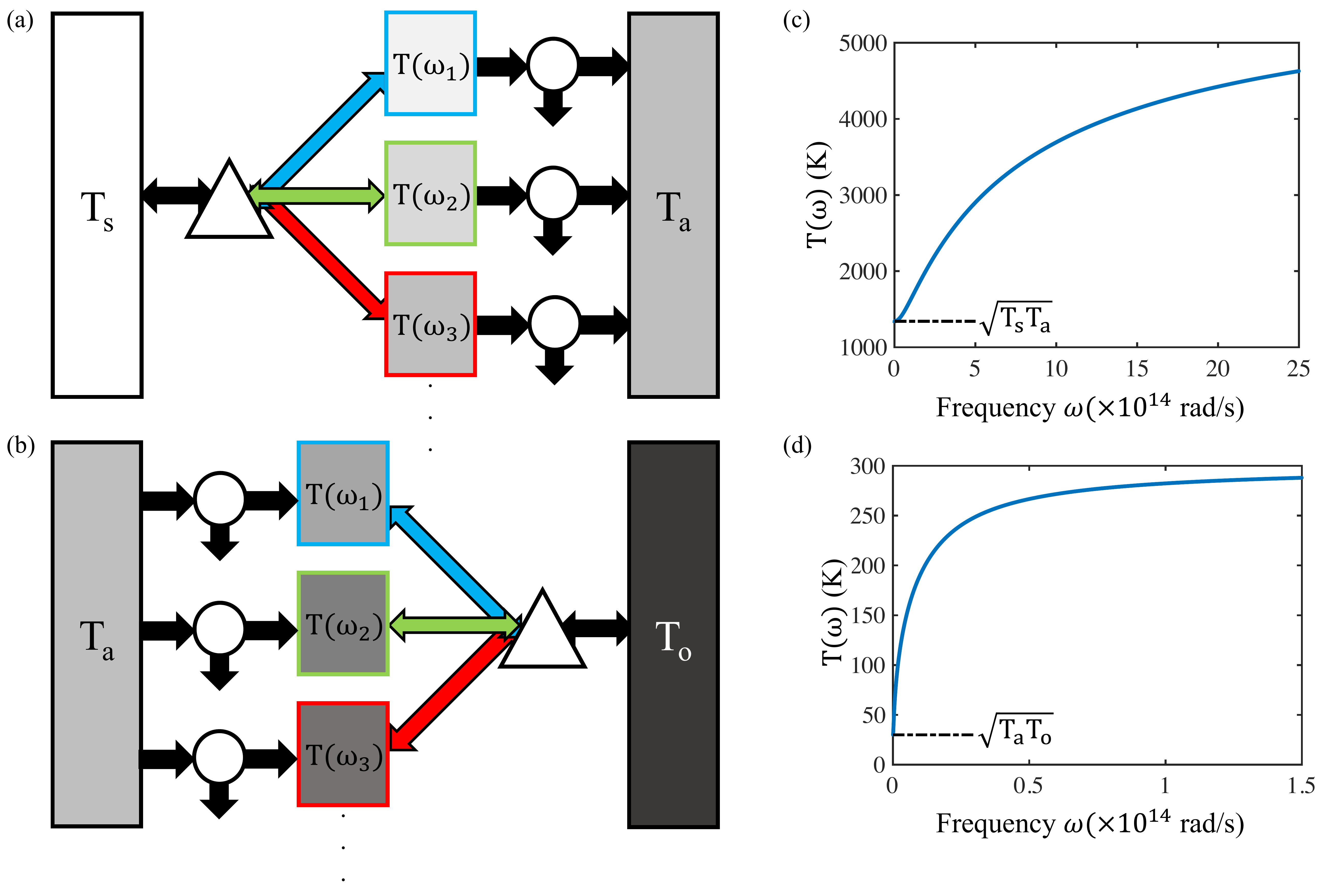}
\caption{Setups to obtain the multi-color limit of energy harvesting under (a) positive illumination, and (b) negative illumination, with the underlying color scheme similar to Fig. \ref{SingleEmitterFigure}. In addition, the colored arrows are indicative of radiative heat exchange at the respective frequencies, and rectangles with colored outlines represent absorbers/emitters that are spectrally selective, with unity absorptivity at the indicated frequency, and zero everywhere else. (c) and (d) Dependence of optimal intermediate temperature on frequency for the setups in (a) and (b), respectively. In both cases, for low frequencies, the optimal intermediate temperature converges to the Chambadal-Novikov-Curzon-Ahlborn limits of $\sqrt{T_sT_a}$ and $\sqrt{T_aT_o}$ respectively.}
\label{MultiEmitterFigure}
\end{figure}
For solar energy conversion, it is known that efficiences higher than that from the setup of Fig. \ref{SingleEmitterFigure}(a) are possible through the setup shown in Fig. \ref{MultiEmitterFigure}(a). In this setup, sunlight first passes through a spectral separator, which separates it into different spectral components. Each spectral component is then sent to heat up a `colored' absorber, which has unity absorptivity at that particular spectral component and zero absorptivity everywhere else. A Carnot engine then operates between each such absorber and the ambient to extract work. Here, the temperature of each absorber can be independently optimized, resulting in an efficiency higher than the blackbody limit. The optimal temperature of absorbers, labelled as $T(\omega)$ for the absorber operating at frequency $\omega$, is shown in Fig. \ref{MultiEmitterFigure}(c). It depicts a monotonic dependence on frequency, i.e., an absorber operating at a higher frequency has a higher optimal temperature. Having obtained $T(\omega)$, the maximum work extracted from the setup of Fig. \ref{MultiEmitterFigure}(a) is
\begin{equation}
\dot{W} = \int_0^\infty d\omega \Big(1-\frac{T_a}{T(\omega)}\Big)\frac{\omega^2}{4\pi^2c^2}\big[\Theta(\omega,T_s)-\Theta(\omega,T(\omega))\big]\label{multisolarstep1}
\end{equation}
where $\Theta(\omega,T) = \hbar\omega/(\exp(\hbar\omega/kT)-1)$. Using this procedure, the multi-color limit on the efficiency of solar energy conversion, $\eta = \dot{W}/\sigma T_s^4$, is 86.8\%, a value slightly higher than the blackbody limit \citep{greenbook}. \\

Based on the duality relation established in Sec. 1, a similar multi-color limit should exist for harvesting outgoing thermal radiation as well. We derive this limit here using a setup obtained by interchanging the radiative transfer elements and Carnot engines of Fig. \ref{MultiEmitterFigure}(a), as shown in Fig. \ref{MultiEmitterFigure}(b). In this setup, `colored' emitters, each with unity emissivity at a particular frequency and zero emissivity elsewhere, lower their temperatures to below the ambient through radiative cooling to the outer space. Carnot engines then operate between the ambient and each such emitter to extract work. Again, the temperature of each emitter can be optimized independently in order to maximize the extracted work. The optimal temperature as a function frequency of the emitter is shown in Fig. \ref{MultiEmitterFigure}(d), where again, a higher frequency emitter has a higher optimal temperature. Using this temperature dependence, the maximum power generated from the setup of Fig. \ref{MultiEmitterFigure}(b) is
\begin{equation}
\dot{W} = \int_0^\infty d\omega\Big(\frac{T_a}{T(\omega)}-1\Big)\frac{\omega^2}{4\pi^2c^2}\big[\Theta(\omega,T(\omega))-\Theta(\omega,T_o)\big] = 55 \ \mathrm{W/m^2}, \label{multitherm}
\end{equation}
which is the multi-color limit of harvesting outgoing thermal radiation, and higher than the blackbody limit of $48.4 \ \mathrm{W/m^2}$ obtained from the setup of Ref. \citep{capasso}. This result was alluded to in Ref. \citep{parthi}, but a detailed derivation was not provided.  \\

In Figs. \ref{MultiEmitterFigure}(c) and \ref{MultiEmitterFigure}(d), we observe that in the low frequency limit, the optimal temperature for the intermediate absorber/emitter approaches $\sqrt{T_sT_a}$ and $\sqrt{T_aT_o}$ for the positive and negative illumination cases, respectively. These results are reminiscent of the optimal converter temperatures in the Chambadal-Novikov-Curzon-Ahlborn limit \citep{novikov, chambadal, curzonahlborn}. In one derivation of this limit, one considers a scenario where an intermediate body at a temperature $T_i$ dissipates heat through conduction to a heat sink at temperature $T_l$. Work is then extracted by a Carnot engine that operates between the intermediate body and a heat source at temperature $T_h$. In this scenario, the maximum work is obtained when $T_i = \sqrt{T_hT_l}$. For the negative illumination case, the low frequency limit corresponds exactly to the scenario above, since in this limit, the net radiative heat flow between the intermediate body and the outer space is proportional to their temperature difference, similar to heat conduction assumed in the derivation of the Chambadal-Novikov-Curzon-Ahlborn limit. \\
\begin{figure}
\centering
\includegraphics[scale=0.6]{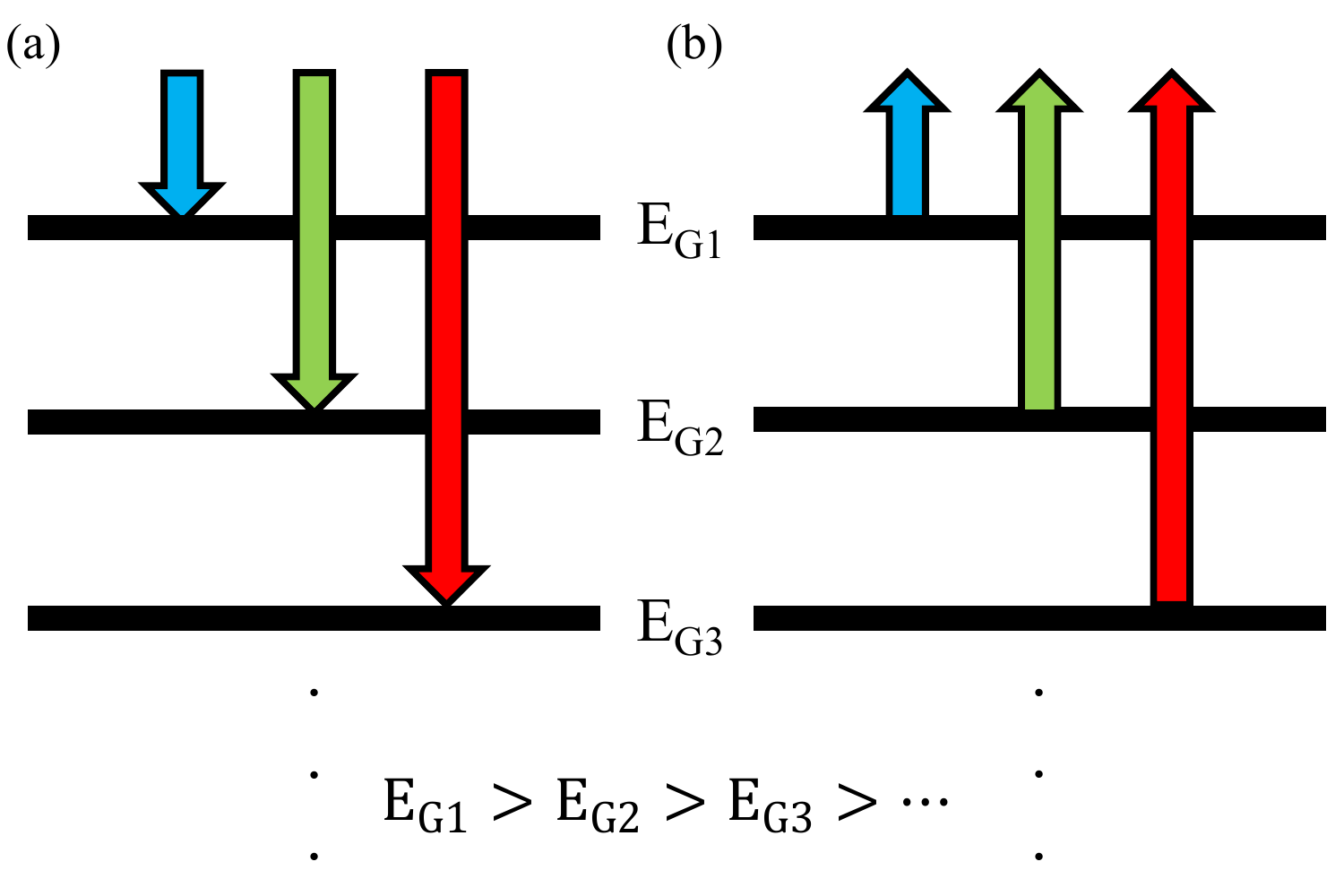}
\caption{Physical implementations of the multi-color limits using stacks of photovoltaic cells with decreasing bandgaps for (a) positive illumination, and (b) negative illumination. The colored arrows are indicative of photon fluxes at the respective frequencies. The attenuation of a frequency at a particular cell indicates its absorption, while its continuation over the cell indicates its transmission. In both cases, the stack is designed such that the cell with the smallest bandgap is nearest to the surface of the earth, which is at the ambient temperature.}
\label{Bandgaps}
\end{figure}


We make a brief comment on the physical implementation of the multicolor limits. For solar energy conversion, instead of using multiple Carnot engines as shown in Fig. \ref{MultiEmitterFigure}(a), one can consider the multi-junction cell configuration in Fig. \ref{Bandgaps}(a), which consists of a stack of semiconductor photovoltaic cells, with the cell near the top of the stack having a larger band gap. When the stack consists of only one cell, the result is the well-known Shockley-Queisser limit \citep{shockleyqueisser} for solar energy conversion, with a maximum efficiency of 40.7\% at $E_g = 1.08$ eV for maximum sunlight concentration, when the cell is at 300 K. The use of more cells in the stack leads to a higher conversion efficiency beyond the Shockley-Queisser limit, as is known in practice with the use of multi-juction solar cells. In this configuration, the multi-color limit of 86.8\% can be reached with the use of an infinite number of cells \citep{marti, browngreen}.\\ 

Similarly, for harvesting outgoing radiation, one can also use the same multi-junction cell configuration as above, as shown in Fig. \ref{Bandgaps}(b), with the cell near the top of the stack having a larger band gap. When the stack consists of only one cell, the resulting maximum power is 54.8 $\mathrm{W/m^2}$, as derived in Ref. \citep{parthi}, which is the Shockley-Queisser limit for harvesting outgoing thermal radiation from a single-junction cell. The multi-color limit of 55 $\mathrm{W/m^2}$ is reached with the use of an infinite number of cells. Unlike the positive illumination case, where there is a substantial improvement by a factor of more than 2 as one goes from a single-junction to a multi-junction cell, in the negative illumination case, the improvement for using a multi-junction cell is far more modest. \\

For solar energy conversion, it was stated that the multi-color limit represents the highest efficiency that can be achieved in a reciprocal system \citep{greenconf}. The same argument should hold for harvesting outgoing radiation as well. However, it is possible to exceed the multi-color limit with the use of non-reciprocal systems, as we will discuss in the next section.  
\section{Landsberg Limit}
\begin{figure}
\centering
\includegraphics[scale=0.42]{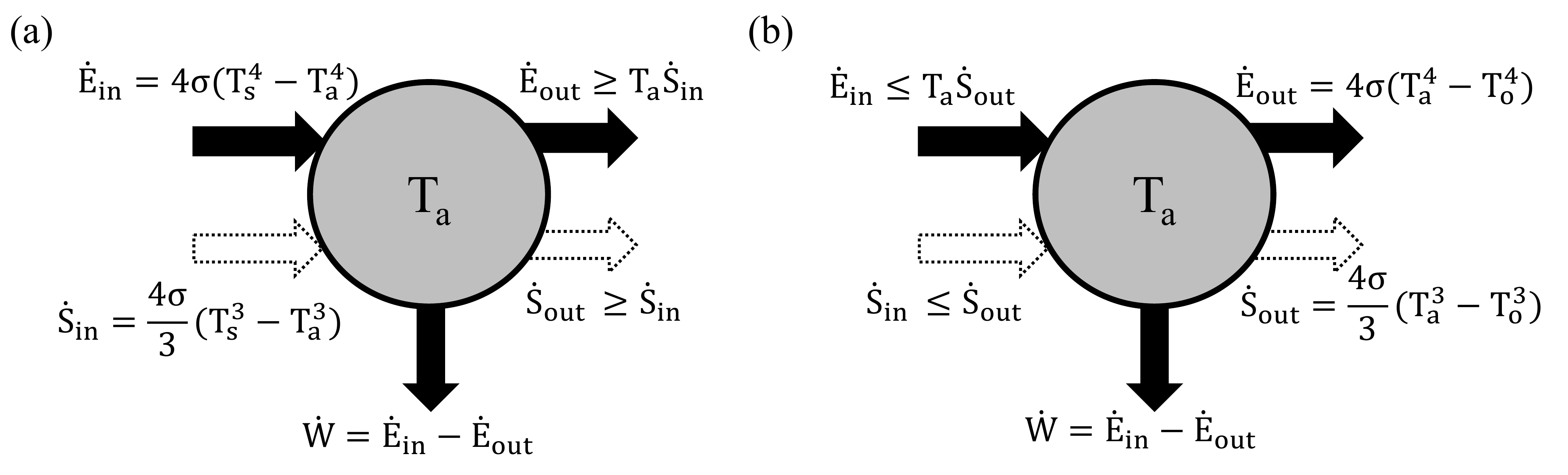}
\caption{Schematics to compute maximum power generated in the Landsberg limit under (a) positive illumination, and (b) negative illumination. The filled black and dotted white arrows represent the \textit{net} energy and entropy fluxes, respectively. The circular disks represent hypothetical engines that can absorb from the heat source and emit to the heat sink, the indicated heat and entropy fluxes. The indicated inequalities are imposed by the second law, which states that the total entropy in a spontaneous process must increase. For positive illumination, the Landsberg limit is obtained by setting the rejected energy flux to its minimum, while for negative illumination, by setting the input energy flux to its maximum.}
\label{LandsbergFigure}
\end{figure}

For solar energy conversion, the ultimate limit is the Landsberg limit \citep{landsberg}, which can be derived from an exergy consideration. Fig. \ref{LandsbergFigure}(a) depicts a schematic to derive the Landsberg limit. We consider a hypothetical engine operating with the sun as the heat source and the ambient as the heat sink, that receives an input heat $\dot{E}_{in}$ equal to the net radiative flux between the sun and the ambient, i.e., $\dot{E}_{in}=\sigma(T_s^4-T_a^4)$, accompanied by the corresponding net entropy flux $\dot{S}_{in} = (4/3)\sigma(T_s^3-T_a^3)$. By the second law, the entropy rejected by the engine to the ambient, $\dot{S}_{out}$, must be at least $\dot{S}_{in}$. In order to find the upper bound on the extraction of work, we assume $\dot{S}_{out} = \dot{S}_{in}$. As a result, the minimum amount of waste heat rejected by the engine to the ambient is $\dot{E}_{out}=T_a\dot{S}_{out} = T_a\dot{S}_{in}$. Therefore, the maximum amount of power that can be generated by this setup is $\dot{W}_\textrm{max}=\dot{E}_{in}-\dot{E}_{out} = \dot{E}_{in} - T_a\dot{S}_{in} = \sigma T_s^4(1 - (4/3)(T_a/T_s) + (1/3)(T_a/T_s)^4)$, corresponding to an efficiency of $\eta = \dot{W}_\textrm{max}/\sigma T_s^4 = 93.3$\% when $T_s = 6000$ K and $T_a=300$ K. This result is the Landserg limit, which represents the ultimate efficiency limit for solar energy harvesting. Here, the expression $\dot{E}_{in} - T_a \dot{S}_{in}$ is commonly referred to as the exergy of incoming heat flow, i.e., the maximum amount of work that can be extracted from an incoming heat flow using an engine at temperature $T_a$. \\

For negative illumination, one can obtain the corresponding ultimate limit of power generation using the schematic depicted in Fig. \ref{LandsbergFigure}(b). This schematic is similar to Fig. \ref{LandsbergFigure}(a), except that we now analyze a hypothetical engine on the hot side instead of the cold side. We assume that the engine has an outgoing heat flux equal to the net radiative flux between the ambient and the cold outer space, i.e., $\dot{E}_{out} = \sigma(T_a^4-T_o^4)$, accompanied by the net entropy flux $\dot{S}_{out} = (4/3)\sigma(T_a^3-T_o^3)$. In order to compute the maximum output power, we are interested in determining the largest heat flux that can be drawn by the engine from the heat source at $T_a$. Since the second law requires the input entropy current $\dot{S}_{in}$ to be \textit{at most} equal to $\dot{S}_{out}$, the maximum heat flux that the engine draws from the heat source is $\dot{E}_{in}=T_a\dot{S}_{in}= T_a\dot{S}_{out}$. Therefore,
\begin{align}
\dot{W}_\textrm{max} &= \dot{E}_{in}-\dot{E}_{out} = T_a\dot{S}_{out} - \dot{E}_{out} \nonumber\\
&=T_a\Big[\frac{4}{3}\sigma(T_a^3-T_o^3)\Big]-\sigma(T_a^4-T_o^4) \label{compareries}
\end{align}
which is equal to 153.1 $\mathrm{W/m^2}$ when $T_a = 300$ K and $T_o = 3$ K. This represents the ultimate limit for harvesting ongoing thermal radiation. Here, similar to the exergy defined for incoming heat flow, one can define $T_a\dot{S}_{out} - \dot{E}_{out}$ as the exergy of the outgoing heat flow. This exergy defines the maximum work that can be extracted from an outgoing heat flow, using an engine that is maintained at a temperature $T_a$. \\
\begin{figure}
\centering
\hskip -0.2in
\includegraphics[scale=0.37]{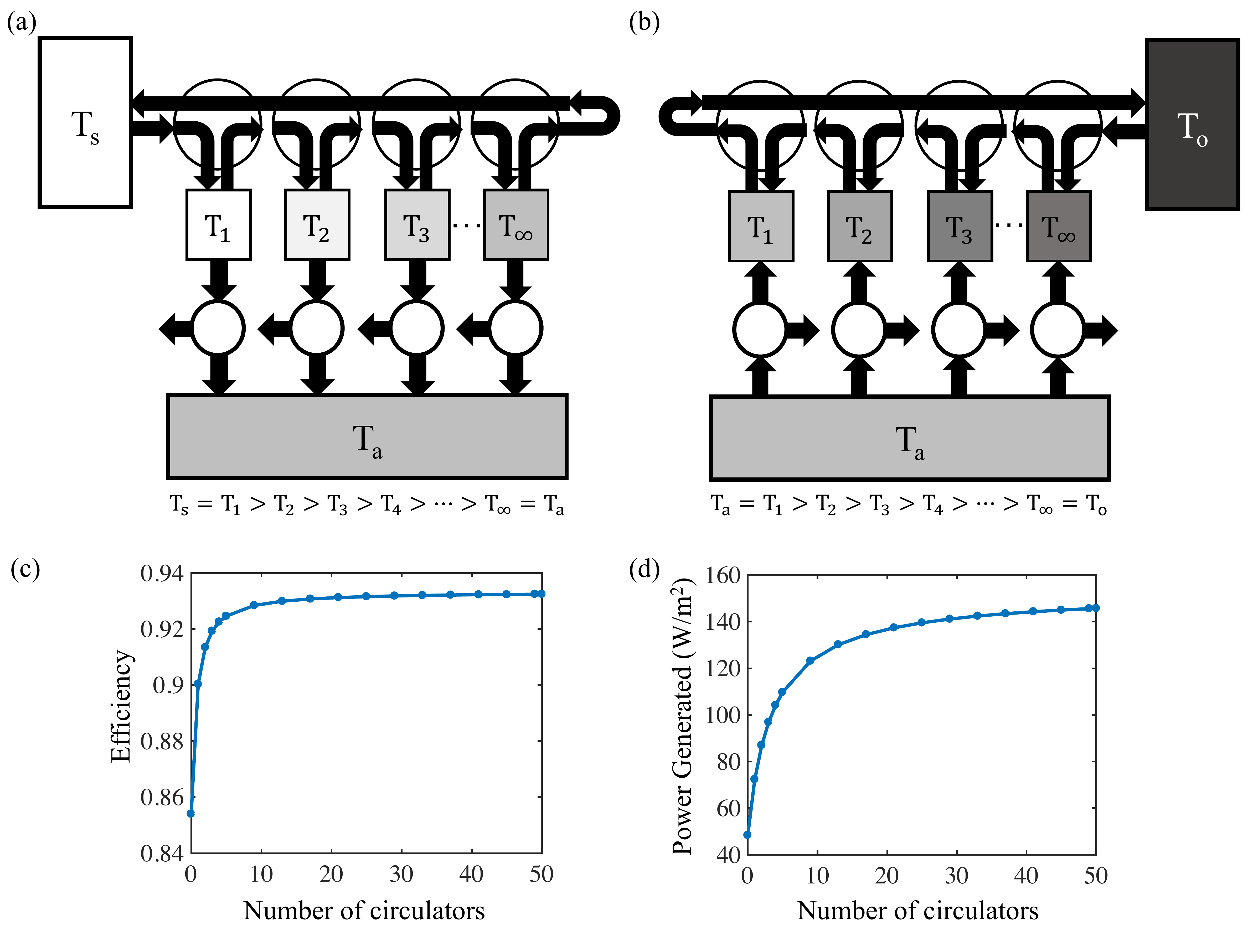}
\caption{Physical realizations of engines to approach the Landsberg limit under (a) positive illumination, and (b) negative illumination. The larger unfilled circles represent circulators, that allow reciprocity breaking to enable unidirectionality of the radiative heat flow. In the limit an infinite number of circulators, these engines attain the Landsberg limits. (c) and (d) Approach to the Landsberg limits as a function of the number of circulators in (a) and (b), respectively. For negative illumination (b, d), with two stages, i.e., a single circulator, the obtained power already exceeds the multi-color limit of 55 $\mathrm{W/m^2}$.}
\label{RiesFigure}
\end{figure}

Attaining the Landsberg limit requires the use of nonreciprocal elements. For the case of positive illumination, Ries \citep{ries, greentime} provided a physical realization for reaching the Landsberg limit using the setup shown in Fig. \ref{RiesFigure}(a). In this setup, the incident sunlight passes through a circulator that directs the light to heat up the first blackbody absorber. A Carnot engine extracts work from the temperature difference between the absorber and the ambient. However, the emission from the first blackbody absorber does not go directly back to the sun, but rather is re-routed by the circulator to heat up a second blackbody absorber. Work is again extracted by a Carnot engine that operates between the second absorber and the ambient. One can construct a multi-stage system in this way, where each stage consists of a blackbody absorber, a circulator, and a Carnot engine. In this system, the last absorber, whose steady state temperature is considerably lower than that of the first absorber, has its blackbody emission radiated back to the sun. As a result, the energy lost in re-radiation in this setup is much smaller as compared to the setup for the blackbody limit of Sec. 1, which corresponds to using only the first absorber stage of the setup in Fig. \ref{RiesFigure}(a). In the limit of an infinite number of intermediate stages, by optimizing the temperature of each absorber, the maximum work extracted corresponds to an efficiency of 93.3\%, which is the same as the Landsberg limit. \\

By the duality relation established in Secs. 1 and 2, it is possible to design a system that achieves the theoretical upper limit of energy harvesting from outgoing thermal radiation. Consider the setup of Fig. \ref{RiesFigure}(b), which is obtained by interchanging the radiative transfer elements and Carnot engines of Fig. \ref{RiesFigure}(a). In this setup, a blackbody emitter radiatively cools to a temperature below the ambient, and a Carnot engine extracts work by operating between the ambient and this blackbody. A thermal circulator directs its emission to a neighboring stage to heat up another blackbody emitter that is radiatively cooling. The latter blackbody emitter consequently attains a temperature intermediate to the previous blackbody and the ambient. Similar to the case of Fig. \ref{RiesFigure}(a), this process can be repeated a large number of times, resulting in a blackbody emitter that is considerably hotter than the emitter receiving an input from the cold outer space. Since the heat current drawn by the Carnot engine from the ambient increases monotonically with the temperature of the intermediate blackbody, this process results in a larger total heat current drawn from the ambient than in the setup for the blackbody limit of Fig. \ref{SingleEmitterFigure}(b), where a single intermediate blackbody emitter is used. We now compute the power generated from this setup in the limit of an infinite number of stages. Consider one of the stages where its blackbody is maintained at temperature $T$. This blackbody receives an input heat flux from its neighboring blackbody at temperature $T-dT$. Consequently, the power generated at this stage by the corresponding Carnot engine is
\begin{align}
\textrm{d}\dot{W} &= \sigma\big[T^4-(T-\textrm{d}T)^4\big]\cdot\Big(\frac{T_a}{T}-1\Big) \nonumber\\
&=4\sigma T^3\textrm{d}T\cdot\Big(\frac{T_a}{T}-1\Big) 
\end{align}
Therefore, the total power generated, obtained by integrating the power over all intermediate temperatures, is
\begin{align}
\dot{W} &= \int_{T_o}^{T_a}4\sigma T^3\textrm{d}T\cdot\Big(\frac{T_a}{T}-1\Big) \nonumber\\
&= \frac{4}{3}\sigma T_a(T_a^3-T_o^3) - \sigma(T_a^4-T_o^4)
\end{align}
which is the same as \eqref{compareries}, and equals 153.1 $\mathrm{W/m^2}$ for $T_a = 300$ K and $T_o = 3$ K. \\

For solar energy harvesting, in Fig. \ref{RiesFigure}(c), we plot the maximum efficiency as we increase the number of stages. By going from one-stage to a two-stage system, the efficiency increases from the blackbody limit of 85.4\% to about 90\% \citep{greenbook, brownas}. (We note that an $N$-stage system requires the use of $N-1$ circulators, since the circulator at the last stage is not strictly necessary.) Similarly, for harvesting outgoing radiation, in Fig. \ref{RiesFigure}(d), we plot the power generated by the negative illumination setup of Fig. \ref{RiesFigure}(b) as a function of the number of circulators used, with all intermediate temperatures optimized to yield the maximum total power. While the ultimate limit can be attained only with an infinite number of stages, a significant gain in power density can be achieved even when a small number of stages is used. For example, the use of two stages instead of one improves the power from 48.4 $\mathrm{W/m^2}$ to 72.3 $\mathrm{W/m^2}$. The two-stage system, therefore, already has a power that is higher than the multi-color limit of Sec. 2. Furthermore, with the use of five stages, one can exceed 100 $\mathrm{W/m^2}$ of power. Therefore, even with a small number stages, and hence a small number of nonreciprocal elements, it is possible to significantly enhance the power extracted from outgoing thermal radiation. \\

\section{Approaching the Carnot efficiency limit in a thermophotovoltaic system}
\begin{figure*}
\centering
\includegraphics[scale=0.36]{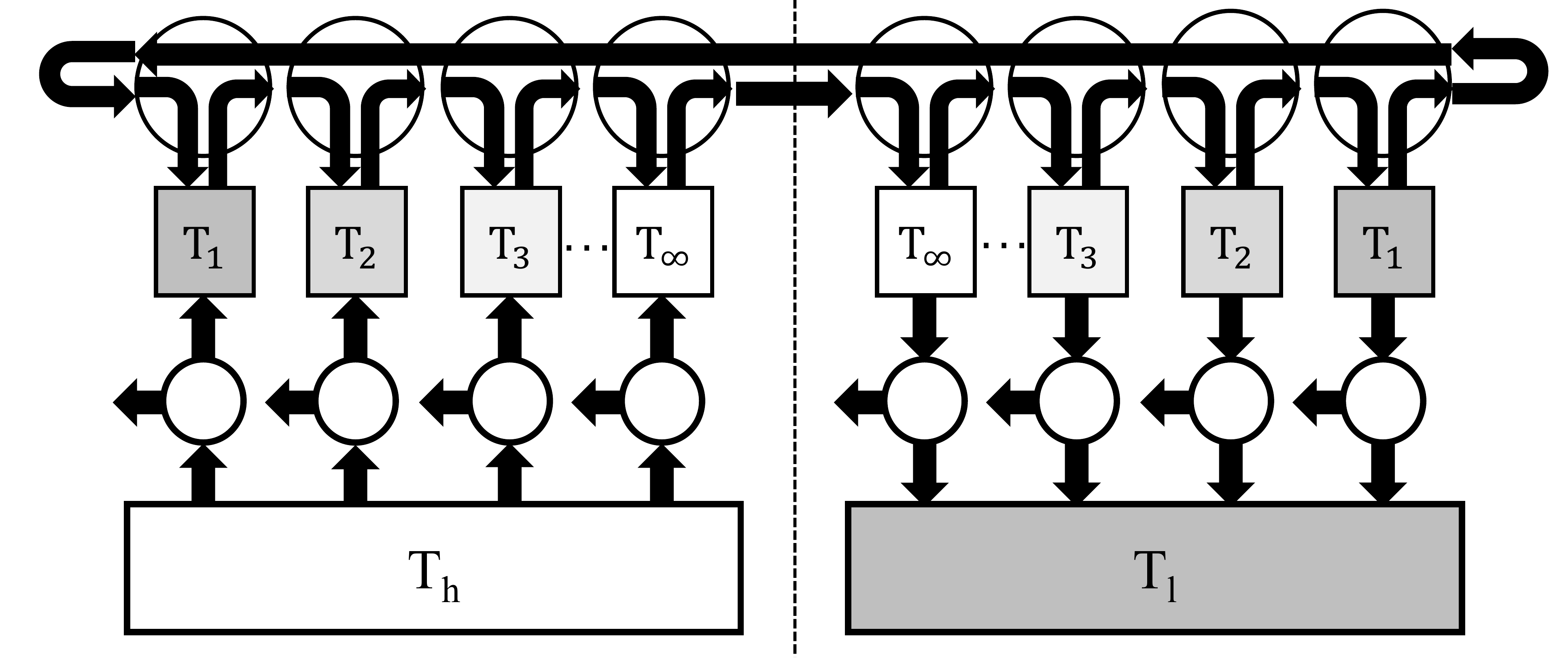}
\caption{Schematic for an engine operating at Carnot efficiency in a thermophotovoltaic system. The setup is obtained by combining the constructions of Fig. \ref{RiesFigure}. The temperatures of the intermediate blackbodies increase from $T_1=T_l$ to $T_{\infty}=T_h$.}
\label{CarnotDiagram}
\end{figure*}
The results from Secs. 1-3 are the main results of the paper, where we elucidate various theoretical limits for harvesting outgoing thermal radiation. We obtained these results by establishing a duality relation between the positive illumination case, where a heat engine operates on the cold side, and the negative illumination case, where the heat engine operates on the hot side. On the other hand, in a thermophotovolatic system, one has access to both the cold and the hot sides. Therefore, one can consider a system as shown in Fig. \ref{CarnotDiagram}, where we place the engines discussed in Fig. \ref{RiesFigure}(a) and \ref{RiesFigure}(b) on the cold and hot sides of the system, respectively. We assume that the temperature of the hot and the cold sides are $T_h$ and $T_l$, respectively. In the system as shown in Fig. \ref{CarnotDiagram}, $T_1 = T_l$ and $T_\infty = T_h$. Therefore, the two sides can connect to each other radiatively without a temperature discontinuity. \\

The engines on either side of the system shown in Fig. \ref{CarnotDiagram} can each extract work that is equal to the exergy appropriate for either incoming or outgoing radiation, for the radiative heat flow defined by an energy flow $\dot{E} = \sigma (T_h^4-T_l^4)$ and an entropy flow $\dot{S} = (4/3)\sigma(T_h^3-T_l^3)$. Therefore, the total work that is extracted from the system is
\begin{equation}
\dot{W} = \big(T_h\dot{S} -\dot{E}\big) + \big(\dot{E}-T_l\dot{S}\big)  = (T_h - T_l)\dot{S}
\end{equation}
Further, one can show from Fig. \ref{LandsbergFigure}(b), with appropriate changes to the temperatures, that the heat provided by the hot side to the engine is $T_h \dot{S}$. Consequently, the efficiency of the system is at the Carnot limit
\begin{equation}
\eta = \frac{\dot{W}}{T_h\dot{S}}= 1 - \frac{T_l}{T_h}
\end{equation}

The system in Fig. \ref{CarnotDiagram} provides an example of a heat engine that operates at the Carnot efficiency limit, but with maximum power output. It is generally accepted in thermodynamics that a finite heat engine operating at the Carnot limit must have only infinitesimal power output. Ref. \citep{casati} argued that a Carnot engine with nonzero power is in principle possible by breaking time-reversal symmetry. Subsequently, Ref. \citep{shiraishi} clarified that in a broad class of systems, including those with broken time-reversal symmetry, a finite engine cannot generate nonzero power at the Carnot limit. Our system certainly breaks time-reversal symmetry since it uses circulators. Moreover, operating exactly at the Carnot limit necessitates the use of an infinite number of stages, and hence a heat engine of infinite size. Thus, our construction here is not in conflict with the existing literature on the subject of power generation at the Carnot limit. On the other hand, our construction does indicate that by breaking time-reversal symmetry, it is possible to achieve both high efficiency and high operating 	power density simultaneously. 

\section{Summary}
In summary, we provide a comprehensive study of various theoretical limits for energy harvesting from outgoing thermal radiation. The results of these limits are summarized in Table 1, where we also compare these limits with the corresponding limits for solar enegy harvesting. Our results suggest that there is significant potential for harvesting outgoing thermal radiation for renewable energy applications. \\

This work was supported by the Department of Energy ``Light-Material Interactions in Energy Conversion” Energy
Frontier Research Center under Grant No. \uppercase{DE-SC}0001293, and the Department of Energy Grant No. DE-FG07-ER46426.

\begin{table}
\centering
\caption{Comparison of positive and negative illumination limits of energy conversion}
\begin{tabular}{lcc}
Limit & Positive Illumination & Negative Illumination \\ & Efficiency & Power\\
1. Shockley-Queisser & 40.7\% & 54.8 $\mathrm{W/m^2}$ \\
2. Blackbody & 85.4\% & 48.4 $\mathrm{W/m^2}$ \\
3. Multi-color & 86.8\% & 55.0 $\mathrm{W/m^2}$  \\
4. Landsberg & 93.3\% & 153.1 $\mathrm{W/m^2}$ \\
\end{tabular}
\end{table}


\end{document}